# Optimal Placement of Data Centers to Support Power Distribution Networks Using Intelligent Algorithms with Economic Indicators

Amin Hajihasani[1], Mahmoud Modaresi[1]

[1]Department of Electrical Engineering, Islamic Azad University, South Tehran Branch, Tehran, Iran

Abstract— Data centers are among the fastest-growing electricity consumers and can impose severe voltage drops and feeder losses when connected to weak distribution networks. This paper formulates a techno-economic siting problem in which each candidate data-center site is mapped to a bus of the distribution network and is assumed to deploy on-site renewable generation and power-electronic interfaces, resulting in a controllable net active-power injection (DG equivalent). A mixed-integer, nonlinear optimization model is developed to jointly select the connection bus and size the DG capacity while respecting network operating limits. The objective combines three normalized terms: (i) active-power losses, (ii) a voltage deviation index capturing profile quality, and (iii) investment cost derived from location-dependent land price and unit DG cost. To address the discrete–continuous search space, an intelligent genetic algorithm (GA) is embedded in a multi-scenario decision framework with adaptive weight tuning: three stakeholder scenarios prioritize losses, voltage quality, or techno-economic balance, and additional balanced scenarios are generated automatically until the optimal bus decision converges. A case study on the IEEE 33-bus radial system demonstrates the effectiveness of the approach. The converged design selects bus 14 with 1.10 MW DG, reducing total losses from 202.67 kW to 129.37 kW (36.2%) while improving the minimum bus voltage to 0.933 p.u. at a moderate investment cost of 1.33 MUSD. The proposed framework provides an interpretable pathway to integrate economic indicators into distribution-aware data-center siting.

## 1. Introduction

Power distribution networks are increasingly required to host new classes of large, spatially concentrated loads. Among them, data centers are particularly impactful because their demand is continuous, scalable, and often co-located with digital infrastructure constraints. Without careful planning, connecting a large data center to a radial feeder can aggravate voltage drops, increase technical losses, and require expensive reinforcements. At the same time, modern data centers are progressively adopting on-site renewable generation, storage, and advanced power conversion, enabling them to behave as controllable prosumers that can support the grid. This motivates a joint planning problem: where should data centers be connected, and what capacity of on-site generation should be installed, so that distribution-network performance and economic indicators are simultaneously satisfied.

This paper presents a distribution-aware techno-economic siting model and an intelligent optimization workflow. The model maps candidate sites to network buses and optimizes a DG-equivalent injection representing the data center's grid-support capability. A genetic algorithm is used to solve the resulting mixed-integer nonlinear problem, and a multi-scenario weighting mechanism is employed to reflect different stakeholder priorities and to promote robust decisions.

## 2. Literature Review and Background

Optimal placement and sizing of distributed generation (DG) has been extensively studied as a means to reduce losses, improve voltage profiles, and enhance operational flexibility of distribution networks. Comprehensive reviews highlight the diversity of objective functions, constraints, and heuristic/metaheuristic solvers used for DG allocation. Genetic algorithms and related population-based methods remain popular because they can handle discrete placement decisions and nonlinear power-flow constraints without requiring convexity. In parallel, the energy footprint and infrastructure cost structure of data centers has motivated research on energy-aware siting and operation, particularly when integrating renewable sources and considering power and cooling overheads. This work builds on these foundations by explicitly coupling an economic site indicator (land cost) with distribution-network metrics within a unified, scenario-weighted GA framework.

## 3. Problem Formulation

Consider a radial distribution network with a set of buses N and branches E. Each candidate data-center site is mapped to one bus of the feeder and is modeled as a controllable net active-power injection (DG equivalent) connected at that bus. The planning variables are: (i) the selected bus index b (with bus 1 as the slack), and (ii) the DG size $P_{DG}$ (kW). Network feasibility and performance are evaluated with a backward/forward sweep power-flow routine, providing bus voltages and total active-power losses.

The optimization minimizes a weighted sum of three normalized terms:
```
F(b,P_DG) = w1·(P_loss/P_loss,0) + w2·(VDI/VDI_0) + w3·(C/C_max) + Penalty
```
where $P_{loss,0}$ and $VDI_0$ are base-case values, and C is the investment cost:
```
C = C_land(b) + c_unit·P_DG.
```
The voltage deviation index is defined as $VDI = \Sigma(1 - |V_i|)^2$ over all buses. A large penalty is applied if any bus voltage violates the limits.

Constraints
- Bus selection: `b ∈ {2,3,…,33}`.
- DG capacity: `P_DG^min ≤ P_DG ≤ P_DG^max`.
- Voltage limits: `0.90 ≤ |V_i| ≤ 1.05` p.u. for all buses.

## 4. Methodology

The solution workflow consists of: (i) distribution-network evaluation via backward/forward sweep load flow, (ii) a GA-based optimizer for the mixed discrete–continuous decision vector, and (iii) a scenario-based weighting loop. Three base scenarios represent distinct priorities: loss minimization, voltage-profile improvement, and techno-economic balance. If the optimal bus does not repeat across scenarios, additional adaptive scenarios are generated using near-uniform weights with small perturbations until the bus decision converges (or a maximum number of attempts is reached). The final design is obtained from the converged scenarios (e.g., by averaging the DG size while keeping the converged bus).

## 4.1 Intelligent Algorithm: Genetic Algorithm

The GA follows the standard cycle of population initialization, fitness evaluation, selection, crossover, mutation, and elitism. Placement is encoded as an integer bus index and sizing as a continuous variable; operators are chosen accordingly to preserve feasibility.

## 4.2 Scenario-based Weighting and Adaptive Convergence

To represent different decision-maker preferences, the optimizer is executed under multiple weight vectors. If the bus decision is inconsistent, additional balanced scenarios are generated automatically until the bus selection converges.

```
Algorithm 1  Multi-scenario GA with adaptive convergence
Input: Network data (branch/bus), economic data C_land, unit cost c_unit,
       base metrics (P_loss,0, VDI_0), GA parameters (N_pop, N_it)
Output: Selected bus b*, DG size P_DG*

1: Define base weight scenarios W_A, W_B, W_C
2: Initialize results list R ← ø, scenario counter s ← 1
3: while not converged and s ≤ S_max do
4:     if s ∈ {A,B,C} then W ← W_s else
5:         W ← normalize( [0.33,0.33,0.33] + small_random_perturbation )
6:     end if
7:     Run GA to minimize F(b,P_DG) under constraints using weights W
8:     Store best solution (b_s, P_DG,s, cost_s, minV_s) into R
9:     if mode of {b_k in R} appears at least twice then
10:        converged ← true;  b* ← mode({b_k})
11:     else
12:        s ← s + 1
13:    end if
14: end while
15: Set P_DG* as the mean P_DG,s among solutions with b_s = b*
```

## 5. Data, Models, and Economic Indicators

The case study uses the IEEE 33-bus radial distribution system (12.66 kV base) with a total active load of 6.015 MW. The maximum DG capacity is set to 3.609 MW (60% of total load) to avoid unrealistically high injections. Economic indicators include a location-dependent land cost, sampled in the range 10–40 kUSD with selected buses marked as urban/expensive (e.g., bus 6 and 30) or rural/cheap (e.g., bus 18), and a unit DG cost of 1200 USD/kW. GA parameters are selected to balance solution quality and runtime.

Table 1. Key parameters used in the study.

| Parameter | Value | Unit |
| --- | --- | --- |
| Base-case losses | 202.67 | kW |
| Base-case VDI | 0.1171 | p.u. |
| Total system load | 6015 | kW |
| Maximum DG capacity | 3609 | kW |

| Unit DG cost | 1200 | USD/kW |
|---|---|---|
| GA population size | 40 | individuals |
| GA iterations | 30 | iterations |

## 6. Experimental Setup and Results

Each scenario is solved with a GA population of 40 individuals and 30 iterations. Each individual encodes a candidate bus index and DG size. Fitness evaluation calls the backward/forward sweep power-flow routine to compute losses and voltages, then evaluates the weighted objective. A penalty is applied for voltage violations. Convergence is declared when the most frequent bus selection appears at least twice across the scenario runs.

Table 2. Scenario definitions and weight vectors.

| Scenario | Priority | Weights [w1,w2,w3] |
|---|---|---|
| A | Loss reduction | [0.80, 0.10, 0.10] |
| B | Voltage quality | [0.10, 0.80, 0.10] |
| C | Techno-economic balance | [0.40, 0.20, 0.40] |
| Adaptive-4 | Balanced (auto) | [0.32, 0.36, 0.32] |
| Adaptive-5 | Balanced (auto) | [0.32, 0.37, 0.31] |

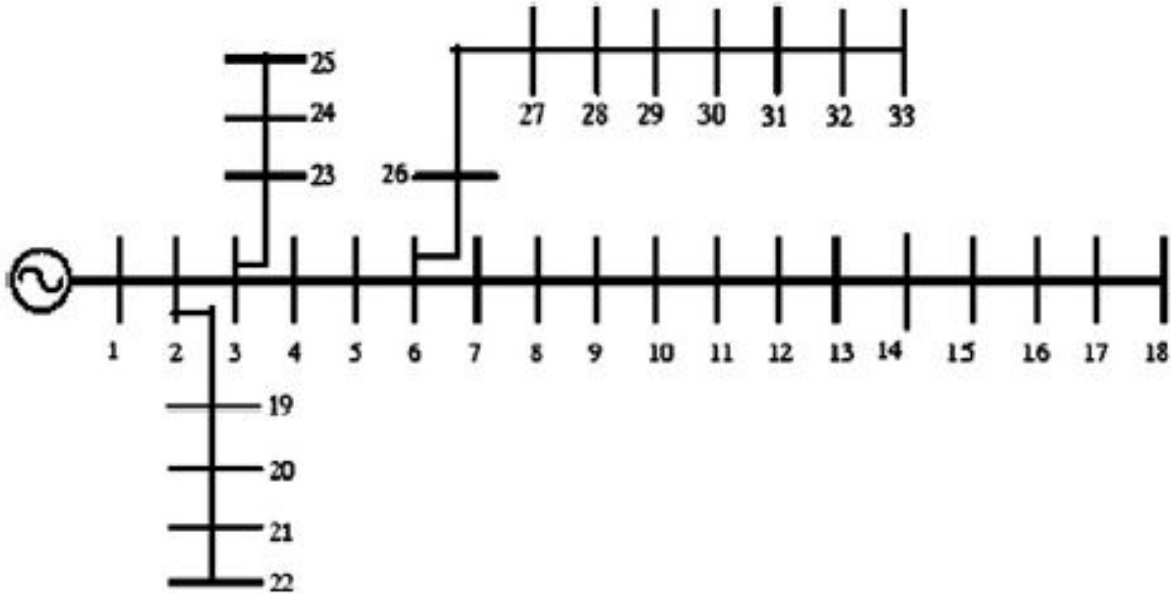

Figure 1. IEEE 33-bus radial distribution test system (case study network).

Table 3. Optimization outcomes across scenarios (values from simulations).

| Case/Scenario | Bus | DG (kW) | Investment (USD) | Loss (kW) | VDI | Min V (p.u.) |
|---|---|---|---|---|---|---|
| Base (no DG) | – | 0 | 0 | 202.67 | 0.1171 | <0.93 |
| A | 7 | 2229.0 | 2,686,542 | 105.66 | 0.0348 | 0.9489 |
| B | 11 | 2229.0 | 2,685,417 | 139.74 | 0.0166 | 0.9489 |
| C | 15 | 755.9 | 922,568 | 137.82 | 0.0566 | 0.9284 |

| Adaptive-4   | 14 | 1082.1 | 1,314,947 | 129.46 | 0.0415 | 0.9331 |
| Adaptive-5   | 14 | 1113.2 | 1,352,253 | 129.31 | 0.0403 | 0.9335 |
| Final (mean) | 14 | 1097.7 | 1,333,600 | 129.37 | 0.0409 | 0.9333 |

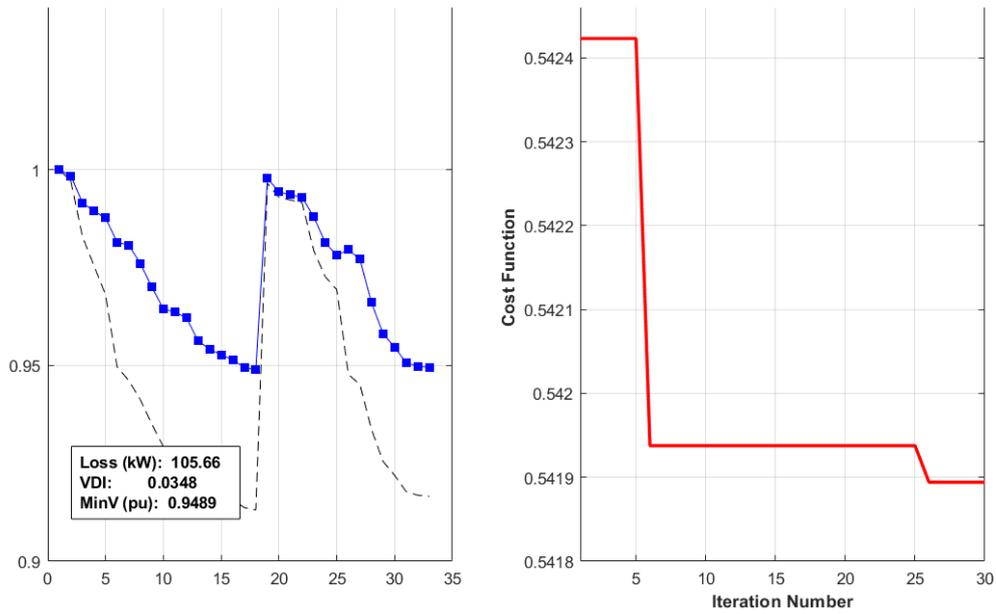

Figure 2. Scenario A (loss-priority): improved voltage profile and GA convergence.

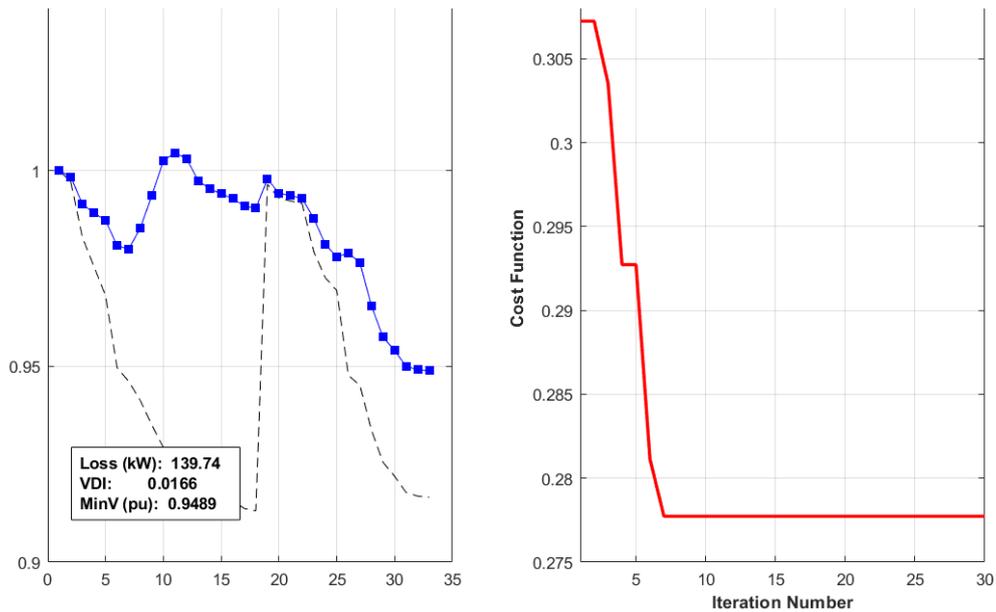

Figure 3. Scenario B (voltage-priority): improved voltage profile and GA convergence.

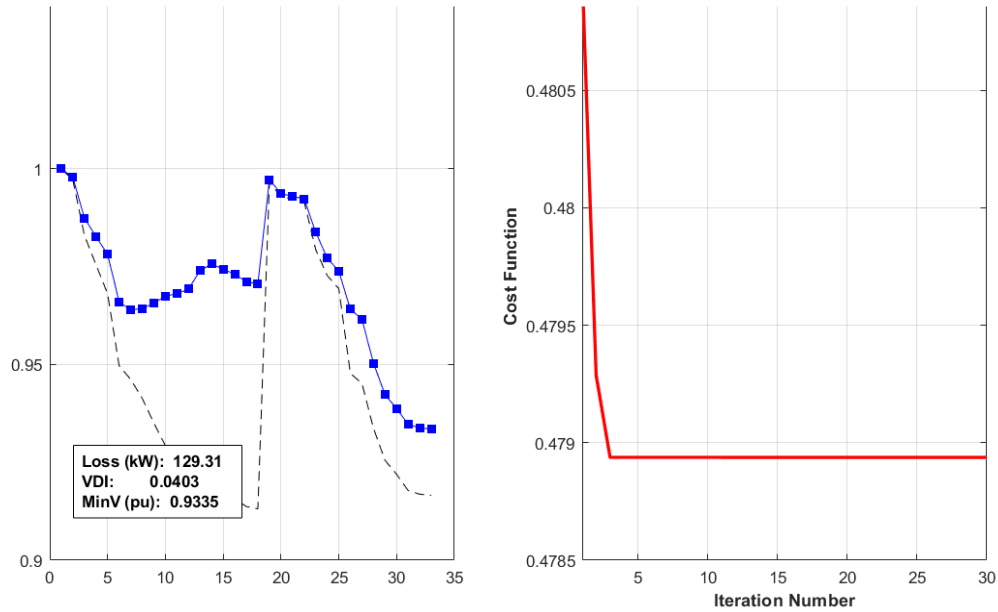

Figure 4. Adaptive converged scenario: balanced voltage and cost performance with converged bus selection.

Table 4. Base-case vs. final design performance.

| Metric | Base case | Final design | Improvement |
| --- | --- | --- | --- |
| Total losses (kW) | 202.67 | 129.37 | 36.2% reduction |
| Minimum voltage (p.u.) | <0.93 | 0.9333 | ~2–3% increase |
| Investment cost (USD) | 0 | 1,333,600 | – |

## 7. Discussion

Scenario A achieves the lowest losses but requires a large investment because it selects a high DG capacity. Scenario B emphasizes voltage quality and yields the best VDI among the evaluated scenarios, again at high cost. Scenario C reduces investment cost substantially, but voltage quality and losses are weaker than in the other scenarios. The adaptive process converges to bus 14, which provides a balanced technical performance with moderate cost, suggesting that this location is a robust compromise under uncertain stakeholder preferences.

From a data-center planning perspective, the converged solution indicates that connecting the facility (with its on-site generation) closer to the mid-feeder can simultaneously relieve downstream voltage drops and reduce upstream losses. Nevertheless, the current model abstracts the data center as a net active-power injection and uses a single operating point. Reliability indices, time-varying load/renewable profiles, and explicit cooling/IT workload constraints should be incorporated in future extensions.

## 8. Conclusion and Future Work

This paper introduced a techno-economic framework for distribution-aware data-center siting using intelligent algorithms. A GA-based optimizer embedded in a multi-scenario and adaptive-weight mechanism was proposed to identify robust siting decisions. On the IEEE 33-bus system, the converged design selected bus 14 with approximately 1.10 MW DG, reducing losses by 36.2% and improving the minimum voltage to 0.933 p.u. at a moderate investment cost. Future work will extend the model to multi-period operation, stochastic renewable profiles, reliability and resilience indices, and quality-of-service constraints specific to data centers.